# Co-author weighting in bibliometric methodology and subfields of a scientific discipline

Lawrence Smolinsky[1] and Aaron J Lercher [2]

[1]*smolinsk@math.lsu.edu*

Department of Mathematics, Louisiana State University, Baton Rouge, LA 70803 (USA)

[2] *alerche1@lsu.edu*

Middleton Library, Louisiana State University, Baton Rouge, LA 70803 (USA)

## Abstract

Collaborative work and co-authorship are fundamental to the advancement of modern science. However, it is not clear how collaboration should be measured in achievement-based metrics. Co-author weighted credit introduces distortions into the bibliometric description of a discipline. It puts great weight on collaboration—not based on the results of collaboration—but purely because of the existence of collaborations. In terms of publication and citation impact, it artificially favors some subdisciplines. In order to understand how credit is given in a co-author weighted system (like the NRC's method), we introduced credit spaces. We include a study of the discipline of physics to illustrate the method. Indicators are introduced to measure the proportion of a credit space awarded to a subfield or a set of authors.

# 1. Introduction

The question of how to award credit for articles with co-authors is important in the research evaluation of researchers and institutions. We argue that some methods for awarding credit favor entire subfields within a discipline.

There are many counting methods to assign credit to co-authors of multi-authored publications. Some methods assign one unit to the whole publication and divide credit among the authors. Some counting methods assign more credit to multi-authored publications. A discussion of the several methods is given by Egghe, Rousseau, and Van Hooydonk (2000). The most generous to co-authors in present use is the "total author counting" method (Egghe et al., 2000, p. 146). This method gives one publication credit to each co-author and so—in effect—an $n$-authored publication is valued at $n$.

Citation counts are often used as a proxy measure of value or impact. In bibliometric discussions of individual articles, the citation counts of article are compared. If an article's value is its citation count, then the value is split among co-authors. In contrast, if each co-author is credited with the full citation count, then the article's value is a multiple of its citation count. We call the second co-author weighted method *authors times citations*. We introduce the notion of credit space to examine the distribution of credit.

This study was partially motivated by the National Research Council (NRC) study of graduate programs in the United States (2011). The NRC is one of the most influential scientific bodies in the world. Co-author weighting is in broad used in research evaluation.

The bibliometric differences between subfields of a discipline may be substantial but are difficult to measure. It requires not only citation data but subfield classification data. Some disciplines have standard ways to classify hundreds of subfields, into which articles may be classified. Yet commonly used citation databases lack such classification structures. In particular, such classification is not available in the Web of Science or SCOPUS.

In this article we present a theoretical analysis of co-author weighting. As an illustration of the theory, we examine subfields of Physics. One straightforward approach for assigning credit would be to measure the percentage of discipline publications that fall in each subfield and the percentage of citations to the discipline that are credited to the subfield. A second method is to follow the co-author weighting methodology for counting publications and citations. The difference between these two methods will be greatest in disciplines where some subfields have many co-authored articles and other subfields have fewer. A brief poster version of this study was first reported in the ISSI2019 conference (Smolinsky & Lercher, 2019).

# 2. US National Research Council

The US National Research Council (NRC) is the principle operating arm of the United States National Academies of Science, Engineering, and Medicine. The NRC periodically evaluates doctorial programs and departments in the US. The most recent evaluation was in 2010 resulting in a report, Assessment of Research Doctorate Programs (National Research Council, 2011).

The report gathered an "unparalleled dataset" and "It enables university faculty, administrators, and funders to compare, evaluate and improve programs" (National Research Council. 2011, p. ix). It "can be used to assess the quality and effectiveness of doctoral programs based on measures important to faculty, students, administrators, funders, and other stakeholders." (National Research Council. 2011, p. 1). The data is meant to influence academic departments. If the report influences the structure of the individual departments in the US, then the report may influence the structure of whole subject fields.

In the sciences, the NRC reported data based on twenty variables. The NRC writes, "Although there was some variation in the faculty responses, they were generally in agreement that publications and citations were the most important factors in program quality." (National Research Council, 2009, p. 12). Of the twenty variables, seven are related to the department faculty (Table 1).

**Table 1**
*NRC assessment of doctorate programs faculty variables*

| Variable |
| --- |
| Publications per allocated faculty |
| Cites per publication |
| Grants per allocated faculty |
| Percent faculty interdisciplinary |
| Percent non-Asian minority faculty |
| Percent female faculty |
| Awards per allocated faculty |

The NRC used 2000-2006 as the period in which to count accumulation of citations to articles published between 1981-2006.

The meaning of the variables labeled *publications per allocated faculty* and *cites per publication* should be understood on a computational level. Articles are credited for authorships and then adjusted for the size of the faculty and reported as annualized rates. For citations, the NRC (2011) described the process using 2003 as an example:

> … the number of allocated citations for a faculty member in 2003 is found by taking the 2003 citations to that faculty member's publications between 1981 and 2003. These counts are summed over the entire faculty in the program and divided by the sum of the allocated publications to the program in 2003. (p. 241)

If an article has two co-authors in two different departments or universities, then each author and department is credited with one publication. If the two co-authors are in the same department, then each author is credited with one article and the department is credited with two articles. Results are then normalized by the size of the department. We follow Egghe et al. (2000) and call this method, *total author counting*. It is consistently used in NRC evaluations.

The situation is slightly more subtle for citations. The full collection of citations to a paper is credited to each co-author using *authors times citations* credit. Suppose an article with two co-authors received two citations. If the two co-authors are in the same department, then the department is only credited with 2 cites/pub since the variable is per publication and the one article was counted twice to the department using *total author counting*. However, if the two co-authors are in different departments or universities, then each department is credited with 2 *cites per publication*.

## 3. Subfield influence

What is the power or influence of one subfield on a discipline? There are likely many different elements to the notions of influence and power. These may include the investment in the subfield in terms of government funding and faculty hiring; the perceived impact on society through press coverage and industrial patents; and the opinions of scientists of relative importance of subfields. Bibliometric data represent one vector of influence. The relationships between the various aspects of scientific influence are likely synergistic. Bibliometric data and grant funding will influence hiring decisions. Discipline composition of the faculty will influence decisions on funding, publication, and hiring—or likely anything that depends on scientific review by committees of faculty members or discipline scientists.

### 3.1 Simple example

We may scientometrically examine these aspects, but methodology decisions shape the results. Consider two highly cited articles. One article has a single author, and a second article has three authors. These articles have quite different effects under co-author weighting where each co-author receives credit for all citations. The single authored article results in one scientist receiving an increase in his influence and prestige in his institution and discipline. The article with three co-authors may result in three influential advocates for their subfield and scientists in their subfield. The single authored article results only in one advocate.

A scientist who authored highly cited papers may have an advantage in applying for funding or for a tenure-track faculty position over a scientist who is not highly cited. Again, the single authored paper results in one advantaged scientist, but the three co-authored paper results in three advantaged scientists.

A US university department that looks to improve its status in the NRC rankings may prefer to hire candidates that will raise their NRC variable count of *cites per publication* and *publications per allocated faculty* counts. The co-authored shared citations and publications improve the profile of all departments that share these citations. The actual contribution of the co-author is not relevant.

### 3.2 Previous studies

There are many ways one might scientometrically examine the question of subfield influence or its individual aspects. Smolinsky and Lercher (2012) studied awards, hiring patterns in prestigious departments, and citations in mathematics subfields. Subfields in medicine (Van Eck, et al. 2013), social science (Walters 2014), computer science (Ibanez, Bielza and Larranaga 2013; Zhu and Yan 2015; Qian et al. 2017) have been studied bibliometrically. Shen et al. (2016) apply a citation input-output analysis to study the influence of physics subfields. Dong et al. (2017) include other factors, including number of co-authors per article and number of publications in subfields of mathematics, physics, and economics. Other factors might include grant funding and faculty surveys.

## 4. Publication and citation spaces

We construct credit spaces for publications and citations. Each point in a credit space is assigned to an individual author. The size of a credit space determines the total amount of credit that is to be allocated.

We use P for publication spaces and C for citation spaces. Let $A$ be a set of articles. We also write $P_A$ for this set. The set $P_A$ is the credit space of publications with one point for each article. The number of articles in $A$ is

(1) $\qquad |P_A|= \sum_{a \in A} 1,$

where the symbol $|S|$ means the number of points in the set S or the size of S.

Let the number of co-authors of an article $a \in A$ be $c_a$ and the number of citations to $a$ be $v_a$. The credit space of authorships for $A$ is denoted $cwP_A$ with a point in $cwP_A$ being an ordered pair of an article and a co-author to that article. The symbol includes cw for co-author weighted. The total number of authorships included in articles in $A$ is

(2) $\qquad |cwP_A|= \sum_{a \in A} c_a.$

The credit space of citations to the articles in $A$ is $C_A$. Each point in $C_A$ is an ordered pair of an article and a citation to the article. There is one point in $C_A$ for each citation to an article in $A$. The total number of citations to articles in $A$ is

(3) $\qquad |C_A|= \sum_{a \in A} v_a.$

The credit space of the co-author weighted citations to the articles in $A$ is $cwC_A$. Each point in $cwC_A$ is an ordered triple of an article, an author of the article, and a citation to the article.

(4) $\qquad |cwC_A|= \sum_{a \in A} v_a\, c_a.$

*Total author counting* for publication credit assigns each point in cwP to an individual the author in the ordered pair (by projection). Likewise, *authors times citations* credit is awarded by assigning each point in $cwC_A$ to an individual scientist, the author in the ordered triple. The amount of total credit awarded in the NRC weighting is given by formulas (2) and (4), rather than (1) and (3).

## 5. Physics

To illustrate the theory, we examine how the credit is split among some groups of subfields of physics as reflected in the Physical Review Journals' classification. Specifically, we compare the fraction of credit received by subfields in the two crediting methods.

### 5.1 Data

The NRC data was not available at the level of detail required, and the NRC did not classify articles by subfields. We use the articles in the American Physical Society's Physical Review journals that are not online-only (American Physical Society 2019). Information is summarized in Table 2. The Physical Review classifies articles into groups of subfields and so our data reflects these groupings rather than individual subfields. We note that applied Physics is not included as Physical Review Applied is one of five online-only Physical Review journals. Physics is also a discipline with many collaborations. It was documented by Jiann-wien Hsu and Ding-wei Huang (2009), who studied the distribution of co-authors in the Physical Review journals and Physical Review Letters.

In order to get an account of the current situation, publication counts were taken from 2013 and allowed the six-year accumulation of citations until January 2019. It gives a significant collection of nearly 15 thousand articles. Nevertheless, there were over 100 thousand articles in physics categories (outside of applied) on the Web of Science in 2013.

**Table 2**
*Physical Review Journals (PR) and their coverage, which reflects the subfields measured*

| PR | Coverage |
|----|----------|
| A | Atomic, molecular, & optical physics and quantum information |
| B | Condensed matter & materials physics |
| C | Areas of experimental & theoretical nuclear physics |
| D | Elementary particle physics, field theory, gravitation, & cosmology |
| E | Statistical, nonlinear, biological, & soft matter physics |

PR = Physical Review journal

## 5.2 Method

The notation is that the set of articles will be $A$ = PR, PRA, PRB, PRC, PRD, or PRE, where PR= Physical Review articles for the year 2013, PRA=Physical Review A articles for the year 2013, etc. Take for example, Physical Review A. The credit space for publications is $P_{PRA}$ and the co-author weighted credit space (i.e., using total author counting) is $cwP_{PRA}$.

The credit space for all of the totality of Physical Review journals is the disjoint union of the credit in individual Physical Review journals:

$P_{PR} = P_{PRA} \cup P_{PRB} \cup P_{PRC} \cup P_{PRD} \cup P_{PRE}$, and

$cwP_{PR} = cwP_{PRA} \cup cwP_{PRB} \cup cwP_{PRC} \cup cwP_{PRD} \cup cwP_{PRE}$.

The total amount of credit awarded to a journal or field is different for the two methodologies and $|P_{PR}| < |cwP_{PR}|$. In this illustration, we will not look at the total credit, but the proportion of credit to a subfield. To be concrete, consider the portion of credit awarded to the subfields of atomic, molecular, & optical physics and quantum information, which is the portion of credit awarded to Physical Review A from the total credit in all Physical Review journals. In the case of a straight publication count, it is

$$AP = |P_{PRA}|/|P_{PR}|.$$

We use AP for article proportion and WAP for weighted article proportion for each Physical Review journal. For credit awarded by co-author weighting, it is

$$WAP = |cwP_{PRA}|/|cwP_{PR}|.$$

To compare the two, we will look at the ratio WAP/AP. For example, if this ratio were 2, then the portion of total credit would be twice as much under total author counting as article counting and if the ratio were ½, then portion would be only half as much.

A similar analysis applies to citation counts. Again, take for example Physical Review A. The credit space of citations to the articles Physical Review A is $C_{PRA}$ and the co-author weighted credit space (i.e., using *authors times citations*) is $cwC_{PRA}$. We use CP for citation proportion and WCP for weighted citation proportion. The proportion of credit to a subfield in terms of citation counts to articles is

$$CP = |C_{PRA}|/|C_{PR}|,$$

and the proportion of credit to a subfield using authors times citations is

$$WCP = |cwC_{PRA}|/|cwC_{PR}|.$$

To compare the two, we will look at the ratio WCP/CP for each Physical Review journal.

**5.3 Results**

Tables 3 and 4 show the credit given to each group of subfields according to each method. The column PR gives the group of subfields specified in Table 2. The column AP in Table 3 gives the fraction of all Physics publication credit that occurs in the subfield under the article counting method; the column WAP give the fraction of all Physics credit that occurs in the subfield under the total author method. The ratio WAP/AP gives the comparison.

Similarly, in Table 4, the column CP in Table 3 gives the fraction of all Physics citation credit that occurs in the subfield under the citation counting method; the column WCP give the fraction of all Physics citation credit that occurs in the subfield under the authors times citation method. The ratio WAP/AP gives the comparison.

**Table 3**
*Comparison of credit to subfields in the Physical Review Journals for publications by two methods of calculation*

| PR | AP | WAP | Ratio |
|----|-------|-------|-------|
| A | 0.195 | 0.054 | 0.277 |
| B | 0.331 | 0.122 | 0.369 |
| C | 0.077 | 0.103 | 1.338 |
| D | 0.224 | 0.680 | 3.036 |
| E | 0.173 | 0.041 | 0.237 |

PR = Physical Review journal
AP = article proportion
WAP = weighted article proportion

**Table 4**
*Comparison of credit to subfields in the Physical Review Journals for citations by two methods of calculation*

| PR | CP | WCP | Ratio |
|----|-------|-------|-------|
| A | 0.158 | 0.038 | 0.241 |
| B | 0.384 | 0.120 | 0.313 |
| C | 0.079 | 0.262 | 3.316 |
| D | 0.262 | 0.555 | 2.118 |
| E | 0.117 | 0.025 | 0.214 |

PR = Physical Review journal
CP = citation proportion
CAP = weighted citation proportion

Some subfields profit and others lose from co-author weighting of publications and citations. The ratios in Tables 3 and 4 show the extent to which credit increases or decreases to subfields. Figure 1 shows how citation credit is split among subfields using a straight count of citations and the NRC method of co-author weighted counts.

**Figure 1**
*Two citation credit spaces in physics using the Physics Review Journals.*

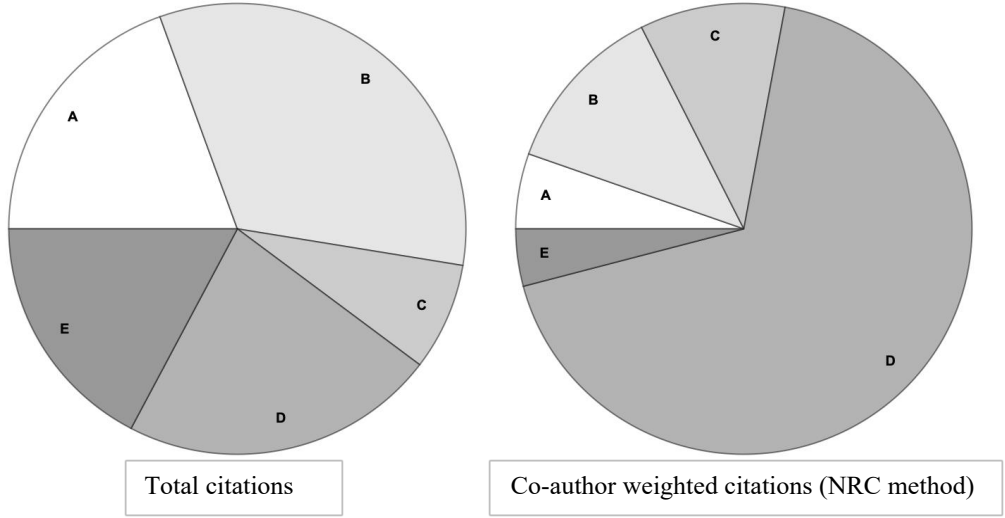

|  |  |
|---|---|
| Total citations | Co-author weighted citations (NRC method) |

Note. Letter denotes area defined by corresponding Physical Review Journal.

## 6. Indicators

The theory and method of sections 4 and 5 may be organized into indicators and expanded. Let $A \subset U$, where $A$ and $U$ are sets of articles, then

$$\text{AP}(A, U) = |P_A|/|P_U| \text{ and WAP}(A, U) = |\text{cwP}_A|/|\text{cwP}_U|$$
$$\text{CP}(A, U) = |C_A|/|C_U| \text{ and WCP}(A, U) = |\text{cwC}_A|/|\text{cwC}_U|$$

Furthermore, one may define WAP($A$, $U$) and WCP($A$, $U$) for any co-author weighted method. Two common examples not previously mentioned are "first author counting" (Egghe et al., 2000; Cole & Cole, 1973) and "fractional author counting" (Egghe et al., 2000; Price 1981). Under both of these crediting methods the weighted counting reduces to the non-weighted counting. This result is reasonable since both these counting methods incorporate corrections for the number of co-authors. One can use more elaborate crediting methods that allow for co-author credit depending on order. They would require adjustments to Eqs 1–4.

Another variation would be to measure the credit proportion to a certain group of authors $G$, which may be a single research group, institution, country, etc. The set of articles $A \subset U$ is then articles in $U$ published with at least one co-author in $G$. Let $c_{aG} =$ number of co-authors of an article $a \in U$ which are in the group $G$. Furthermore, let $\text{gcwP}_A \subset \text{cwP}_A$, and $\text{gcwC}_A \subset \text{cwC}_A$, be only those subsets of tuples with the co-author entry in $G$. These capture only the data for authors in the group $G$. We use g and G for group. It follows that $|\text{gcwP}_A| = \sum_{a \in A} c_{aG}$ and $|\text{gcwC}_A| = \sum_{a \in A} v_a\, c_{aG}$. The indicators that capture the proportion of the weighted credit space due to the authors in $G$ are

$$\text{GWAP}(A, U) = |\text{gcwP}_A|/|\text{cwP}_U|$$
$$\text{GWCP}(A, U) = |\text{gcwC}_A|/|\text{cwC}_U|$$

One can use the group weighted article and citation proportion to measure the proportion of the collaborative effort due to $G$ or compare the values to AP($A$, $U$) and CP(A, U) to contrast with the credit proportions unweighted by co-authors.

### 7. Conclusion

Collaborative work and co-authorship are fundamental to the advancement of modern science. However, it is not clear how collaboration should be measured in achievement-based metrics. Co-author weighted credit introduces distortions into the bibliometric description of a discipline. It puts great weight on collaboration—not based on the results of collaboration—but purely because of the existence of collaborations. In terms of publication and citation impact, it artificially favors some subdisciplines. In order to understand how credit is given in a co-author weighted system (like the NRC's method), we introduced credit spaces. One can explicitly observe the effects in subdisciplines with indicators AP, WAP, AC, and WAC. It is also desirable to understand the portion of total credit that can be attributed to various institutions, regions, and other groups of authors. The indicators GWAP and GWAC allow for this measurement and to judge value in co-author weighted credit space.

One possible objection to this analysis is that the NRC did not intend its data to be used in this way for subfield analysis. But the same can be said for using bibliometric databases to gather citation counts to evaluate researchers. The Scientific Citation Index (SCCI) was not primarily intended to be used for evaluating individual scientists and institutions nor evaluating impact, but it is routinely used for these purposes—including by the NRC. SCCI was a tool for the scientific community to find connected research and guide researchers. It was inspired by a legal index to court cases. Garfield wrote, "The legal 'citator' system provided a model of how citation index could be organized to function as an effective search tool" (1979, p. 7). Garfield proposed that a citation index could be used to evaluate journals, and for historical evaluation of "the significance of a particular work and its impact on the literature and thinking of the period." (1955, 109) But the primary intended purpose of the Scientific Citation Index was to assist scientists in their use of scientific literature.

One may argue that the co-author list may indirectly reflect other legitimate measures of a subfield such as the number of researchers or funding. However, the measures should not confound different aspects. Grants per faculty member reflects the size of the funded pool. The number of co-authors may or may not reflect the number of researchers, but it may inflate the importance of each co-author by giving each a citation credit.

Collaboration has become valued for its own sake and there is evidence that collaboration also impacts traditional bibliometric variables, e.g., there is evidence that co-authored papers receive more citations (Onodera and Yoshikane, 2015). However, co-author weighting may have consequences such as offering arbitrage opportunities for adding authors, e.g., ghost authors (Smolinsky, 2020).

The NRC also constructed weights on the twenty variables using a regression methodology (called R-based) as well as using the faculty survey data (S-based). With the R-based weights the highest weighted three variables in Physics were *Awards per Allocated Faculty*, *Average GRE-Q*, and *Average PhDs 2002 to 2006*. The bibliometric faculty variables became less prominent with citations dropping from second in the S-based median ranking to eleventh in the R-based median ranking of weights. Publications had a less dramatic drop from third to fourth. It could be that citations were ranked less important in the R-based measure because there is no difference between the first author or the tenth author in the authors times citations measure. Alternatively, citations may have ranked high in the S-based measure because faculty members were not completely aware of what they were selecting.

The greatest impediment to further research is the lack of subfield information in standard bibliometric data bases. Sometimes subfield information is contained in classifications from the professional societies, journal collections, or preprint archives, but the difficulty in merging this information makes full indicator calculations difficult. Immediate future research will likely rely on sampling. Examination of the credit received by institutions does not necessarily require subfield information.